\documentclass{PoS}

\title{Supergiant Fast X-ray Transients: interpretation of archival
  INTEGRAL data}

\ShortTitle{Supergiant Fast X-ray Transients: interpretation of archival
  INTEGRAL data}

\author{\speaker{L.~Ducci}$^{ab}$,
  L.~Sidoli$^b$, A.~Paizis$^b$, S.~Mereghetti$^b$\\
  \llap{$^a$} Dipartimento di Fisica e Matematica, Universit\`a degli
  Studi dell'Insubria,\\ 
  Via Valleggio 11, I-22100 Como, Italy \\
  \llap{$^b$} INAF, Istituto di Astrofisica Spaziale e Fisica
  Cosmica,\\
  Via E. Bassini 15, I-20133 Milano, Italy \\

  E-mail: \email{lorenzo@iasf-milano.inaf.it}
}

\abstract{\emph{INTEGRAL} monitoring of the Galactic Plane in the last 5 years
revealed a new subclass of High Mass X-ray Binaries (HMXBs), the
Supergiant Fast X-ray Transients (SFXTs). They display flares lasting
from minutes to hours, with peak luminosity of 
$10^{36}-10^{37} \ erg \ s^{-1}$ and a frequent long term flaring
activity reaching an X-ray luminosity of 
$10^{33} - 10^{34} \ erg \ s^{-1}$, as recently detected by the
\emph{Swift} satellite. The quiescent level is around 
$10^{32} \ erg \ s^{-1}$. We performed a systematic re-analysis of
archival \emph{INTEGRAL} data of four SFXTs: IGR~J16479-4514,
XTE~J1739-302, IGR~J17544-2619, IGR~J18410-0535. 
This led to the discovery of previously unnoticed outbursts from 
IGR~J16479-4514 and IGR~J17544-2619. We discuss these
results in the framework of the different structure of the supergiant
wind proposed to explain the outburst from this new class of sources.}

\FullConference{7th INTEGRAL Workshop\\
		 September 8-11 2008\\
		 Copenhagen, Denmark}

\begin{document}

\section{Introduction}

The \emph{INTEGRAL} monitoring of the Galactic Plane 
led to the discovery of many new
High Mass X-ray Binaries (HMXBs), and in particular 
of a new class of X-ray transients with OB supergiants,
the Supergiant Fast X-ray Transients (SFXTs)
(\cite{Negueruela et al. 2006}; \cite{Sguera et al. 2005}).
SFXTs are characterized by fast X-ray flares 
with peak luminosity of $10^{36} - 10^{37} \ erg \ s^{-1}$, and
a frequent long-term flaring activity with a level of X-ray luminosity
of $10^{33} - 10^{34} \ erg \ s^{-1}$ \cite{Sidoli et al. 2008}.

The main hypotheses proposed to explain the SFXTs behaviour 
are based on the structure of the supergiant wind 
(see Sidoli 2008 for a recent review \cite{Sidoli 2008}):
flares could be due to accretion of clumps in a clumpy spherical wind
(in't Zand 2005 \cite{in't Zand 2005}) 
or to enhancement of accretion when the neutron star
crosses an equatorial wind component 
(Sidoli et al. 2007 \cite{Sidoli et al. 2007}).

Another model involves the presence of magnetars in SFXTs, and the
short outbursts result from transition across the magnetic and
centrifugal barriers (Bozzo et al. 2008 and references therein 
\cite{Bozzo et al. 2008}).

Here we report the results of a systematic analysis of public
archival \emph{INTEGRAL} data of 4 SFXTs: IGR~J16479-4514,
XTE~J1739-302, IGR~J17544-2619, IGR~J18410-0535
from 2003 to 2007.
We compare the observational results on these 4 SFXTs
with the predictions of the spherical clumpy stellar wind model that
we have developed.

\section{Observations, data analysis and results}

\begin{figure}[htbp]
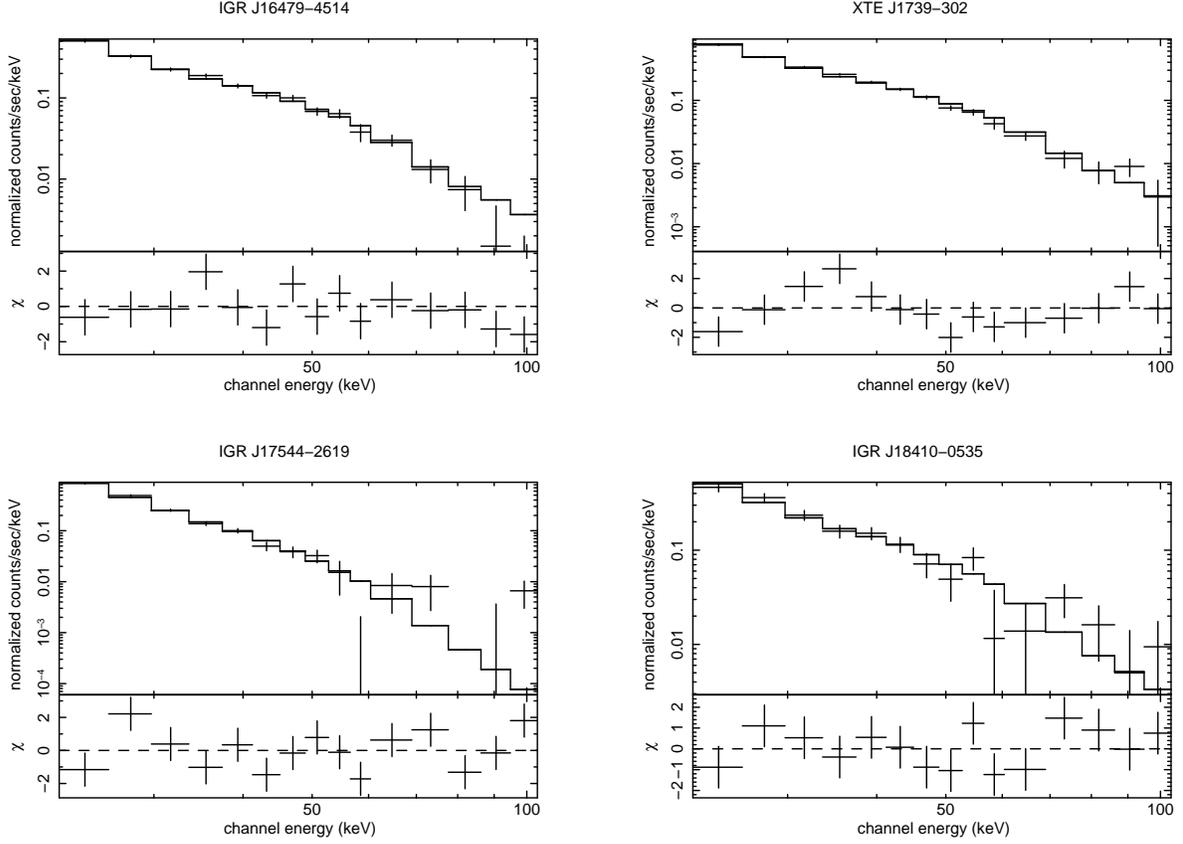

\begin{center}
\begin{tabular}{cc}
\includegraphics[angle=-90, width=8cm]{./ps_files/spettro22_100keV_igrj16479_bremss.ps} &
\includegraphics[angle=-90, width=8cm]{./ps_files/spettro22_100keV_igrj17391_bremss.ps} \\
\includegraphics[angle=-90, width=8cm]{./ps_files/spettro22_100keV_igrj17544_bremss.ps} &
\includegraphics[angle=-90, width=8cm]{./ps_files/spettro22_100keV_igrj18410_bremss.ps} \\
\end{tabular}
\end{center}
\caption{Bremsstrahlung fit and residuals (in units of standard deviations) 
         to the average IBIS/ISGRI spectra of IGR~J16479-4514, 
         XTE~J1739-302, IGR~J17544-2619, IGR~J18410-0535.}
\label{figura_4_spettri}
\end{figure}

Using OSA 7.0, we analysed $7328$ ScWs (IBIS/ISGRI) ranging from
2003 to 2007, corresponding to a total exposure time of $15 \ Ms$
for the 4 SFXTs.
We extracted the lightcurves in the energy range $20-60$~keV 
and we found 31 flares for IGR~J16479-4514, 43 flares for XTE~J1739-302,
16 flares for IGR~J17544-2619 and 3 flares for IGR~J18410-0535.
Besides the outbursts already reported in the literature, we
discovered new flares, listed in Table (\ref{INTEGRAL_observations}).
\begin{table}
\begin{center}
\begin{tabular}{cccc}
\hline
\multicolumn{4}{c}{IGR~J16479-4514} \\
\hline
N. & Start Time                 & Peak flux ($20-60$~keV) & det.\\
   &    (UTC)                   & $erg \ cm^{-2} \ s^{-1}$ & significance \\
\hline
1  & 2003 Feb 02,  19:06  & $5.9 \times 10^{-10}$    & $5.95$\\
2  & 2004 Aug 09,  02:24  & $1.3 \times 10^{-10}$    & $5.46$\\
3  & 2004 Aug 20,  07:26  & $7.0 \times 10^{-10}$    & $5.38$\\
4  & 2004 Aug 20,  12:04  & $1.2 \times 10^{-9}$     & $7.28$\\
5  & 2004 Sep 10,  01:12  & $5.8 \times 10^{-10}$    & $6.04$\\
\hline
\hline
\multicolumn{4}{c}{IGR~J17544-2619} \\
\hline
1  & 2004 Feb 27,  14:16  & $1.0 \times 10^{-9}$    & $6.29$\\
2  & 2006 Sep 20,  10:00  & $2.3 \times 10^{-9}$    & $8.39$\\
\hline
\end{tabular}
\end{center}
\caption{Summary of the new flares of IGR~J16479-4514 and
  IGR~J17544-2619 discovered in this work. 
  The peak flux is calculated on a time interval of 200~s.}
\label{INTEGRAL_observations}
\end{table}
For each outburst, we have extracted an IBIS/ISGRI spectrum in the energy
range $22-100$~keV and we have performed
a fit with different spectral models: power law, bremsstrahlung,
Comptonization model ({\sc comptt} in {\sc xspec}). 
Since there was no evidence of a spectral
difference between the flares, within the uncertainties, we extracted
a total flare spectrum for each of the 4 SFXTs. We obtained the best
fit with a bremsstrahlung model (Figure \ref{figura_4_spettri}), with the parameters 
reported in Table (\ref{tabella_parametri_spettri}).

\begin{table}
\begin{center}
\begin{tabular}{lccc}
\hline
                &$T_{exp}$&  kT       &   $\chi^2_{\nu}$  \\
                &(days)  & (keV)     &   (13 \ d.o.f.)   \\
\hline
IGR~J16479-4514 & 47  & $27.0{+2.1 \atop -1.9}$ & $1.04$ \\
XTE~J1739-302   & 118 & $22.2{+0.9 \atop -1.1}$ & $1.71$ \\
IGR~J17544-2619 & 118 & $10.4{+0.6 \atop -0.6}$ & $1.56$ \\
IGR~J18410-0535 & 35  & $26.4{+7.1 \atop -5.1}$ & $0.94$ \\
\hline
\end{tabular}
\end{center}
\caption{Best fit parameters of the average spectra of 
         IGR~J16479-4514, XTE~J1739-302, IGR~J17544-2619 
         and IGR~J18410-0535 with a bremsstrahlung model (IBIS/ISGRI).
         $T_{exp}$ is the net exposure time.}
\label{tabella_parametri_spettri}
\end{table}

\section{Clumpy stellar winds} \label{Section Clumpy stellar winds}

We developed a spherical clumpy stellar wind model in order to compare
predictions of this model with the behaviour of the SFXTs.
We assumed that the OB type supergiant is surrounded by a clumpy and
spherically symmetric wind, where the clump formation rate
distribution is:
\begin{equation} \label{Npunto}
\dot{N} = k M_{cl}^{-\zeta} \ \ clump \ s^{-1}
\end{equation}
where $M_{cl}$ is the mass of the clump,
and $\dot{N}$ is the rate of clumps with mass
$M_{cl}$ emitted by the star. 
We assumed spherical clumps,
with radii $R_{cl}$, then we
introduce a power law distribution of radii $R_{cl}$:
\begin{equation} \label{distrib R_cl}
\dot{N}_M \propto R_{cl}^{\gamma} \ \ clump \ s^{-1}
\end{equation}
We assumed that the total mass loss rate of the supergiant 
is given by:
\begin{equation} \label{total mass loss rate}
\dot{M}_{tot} = \dot{M}_{cl} + \dot{M}_{wind}
\end{equation}
where $\dot{M}_{cl}$  is the component of mass loss rate 
due to the clumps and $\dot{M}_{wind}$ is the
mass loss rate in the form of the tenuous inter-clump medium.

From equations (\ref{Npunto}) and (\ref{total mass loss rate}) we obtain
that the total mass loss rate of the supergiant, $\dot{M}_{tot}$, is given by:
\begin{equation} \label{Mpunto}
\dot{M}_{tot} = \dot{M}_{wind} + \int_{M_a}^{M_b} \dot{N} dM =
\dot{M}_{wind} + \int_{M_a}^{M_b} k M^{-\zeta} dM
\end{equation}
In equation (\ref{Mpunto}) $M_a$ and $M_b$ are the limits of the
clump mass range. 
Assuming that $\dot{M}_{tot}$ is known,
from Equation (\ref{Mpunto}) we can determine $k$:
\begin{equation} \label{k}
k = \frac{f\dot{M}_{tot}}{\int_{M_a}^{M_b} M^{-\zeta} dM} = \frac{f\dot{M}_{tot}}{\frac{1}{1 - \zeta}(M_b^{1 - \zeta} - M_a^{1 - \zeta})}
\end{equation}
where $f=\dot{M}_{cl} / \dot{M}_{tot}$ is the fraction of wind mass
contained in the clumps.

Clumps are driven radially outward by transfer of momentum from UV photons
to the ions of the wind via absorption or scattering in spectral lines
\cite{Castor et al. 1975}. 
From spectroscopic observations of O stars, Lepine \& Moffat (2008)
\cite{Lepine and Moffat 2008}
suggest that clumps have the same velocity law of a smooth stellar
wind. We can then assume for the clump velocity profile $v(r)$:
\begin{equation} \label{legge_velocita}
v(r) = v_{\infty}\left (1 - 0.9983\frac{R_{OB}}{r} \right )^{\beta}
\end{equation}
where $v_{\infty}$ is the terminal wind speed, $R_{OB}$ is the radius of
the supergiant, $0.9983$ is a dimensionless parameter which ensures
that $v(R_{OB}) \approx 10 \ km \ s^{-1}$, and $\beta=0.8$ is a constant
(\cite{Lamers and Cassinelli 1999}; \cite{Kudritzki et al. 1989}).

\begin{figure}[htbp]
\begin{center}
\includegraphics[width=10cm]{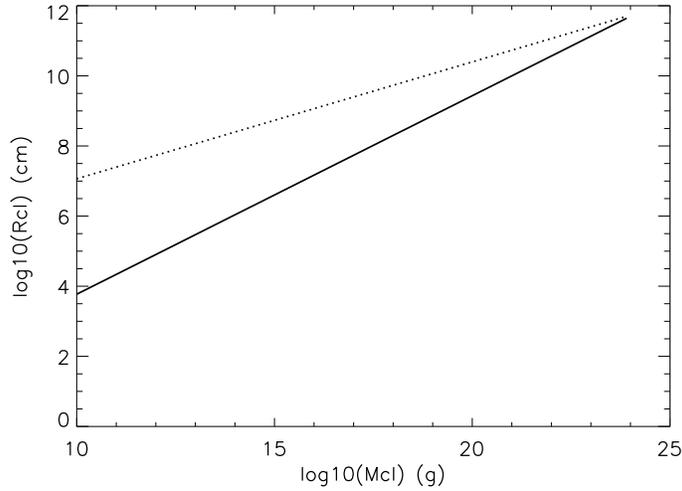}
\end{center}
\caption{Functions of the upper-limit (dotted line) and lower-limit
  (solid line) of the clump radius at $r=R_{OB}$. In order to obtain these
  functions we have assumed the following parameters for the
  supergiant wind: $M_{OB}=30 \ M_{\odot}$, $R_{OB}=23.8 \ R_{\odot}$, 
  $v_{\infty}=1700 \ km \ s^{-1}$, $\beta=0.8$, $v_0=10 \ km \ s^{-1}$,
  $\zeta=2$, $\gamma=-1.5$, 
  $M_a = 5 \times 10^{19} \ g$,
  $M_b= 10^{22} \ g$,
  $\dot{M}_{cl}/\dot{M}_{wind}=0.85$.}
\label{relazione_Mcl_Rcl_Mloss015}
\end{figure}

The clump size is determined by the balance pressure equation.
Following Lucy \& White (1980) \cite{Lucy and White 1980} 
and Howk et al. (2000) \cite{Howk et al. 2000}, 
we find that the clump size increases at larger distances from the
supergiant star (see our derivation in 
Romano et al. (2008a) \cite{Romano et al. 2008a})
with the equation:
\begin{equation} \label{legge_Rcl_r}
R_{cl}(r) = R_{cl}(R_{OB}) \left ( \frac{r^2 v(r)}{R_{OB}^2 v_0} \right )^{1/3}
\end{equation}
where $v_0=v(R_{OB})$ is the initial velocity of the clump at the
surface of the supergiant.
We found, for each mass of the clump, the upper-limit and the
lower-limit for the clump radius by means of two conditions.
For any given mass there is a minimum radius below which 
the clump is optically thick in the UV resonance
lines. Then the gravity dominates over the radiative force due to the
line scattering and the clump tends to fall back onto
the supergiant.
In order to be accreted by the compact object of the binary
system, the clump must  escape from the OB supergiant.
Then, from the theory of radiatively driven stellar wind of Castor et
al. (1975) \cite{Castor et al. 1975}, and the clump model of Howk et
al. (2000) \cite{Howk et al. 2000}, we found for each mass of the
clump the lower-limit for its radius.
With regard to the second condition, the clump is defined as a density
enhancement in the smooth stellar wind.
Then, for each mass of the clump, there exists an upper-limit for the clump
radius: clumps with larger radii would be less dense than the smooth
stellar wind (inter-clump medium), in contrast with the clump
definition.
The functions of the upper-limit and lower-limit for the clump radius
are represented in Figure (\ref{relazione_Mcl_Rcl_Mloss015}).

In conclusion, it is possible to derive the 
properties of the supergiant stellar wind 
assuming the clump mass distribution
(\ref{Npunto}), the clump radii distribution (\ref{distrib R_cl}),
the expansion law of the clump (\ref{legge_Rcl_r}) and the upper-limit
and lower-limit laws for the clump radius 
(see Figure \ref{relazione_Mcl_Rcl_Mloss015}).

\begin{figure}[htbp]
\begin{center}
\begin{tabular}{cc}
\includegraphics[width=8cm]{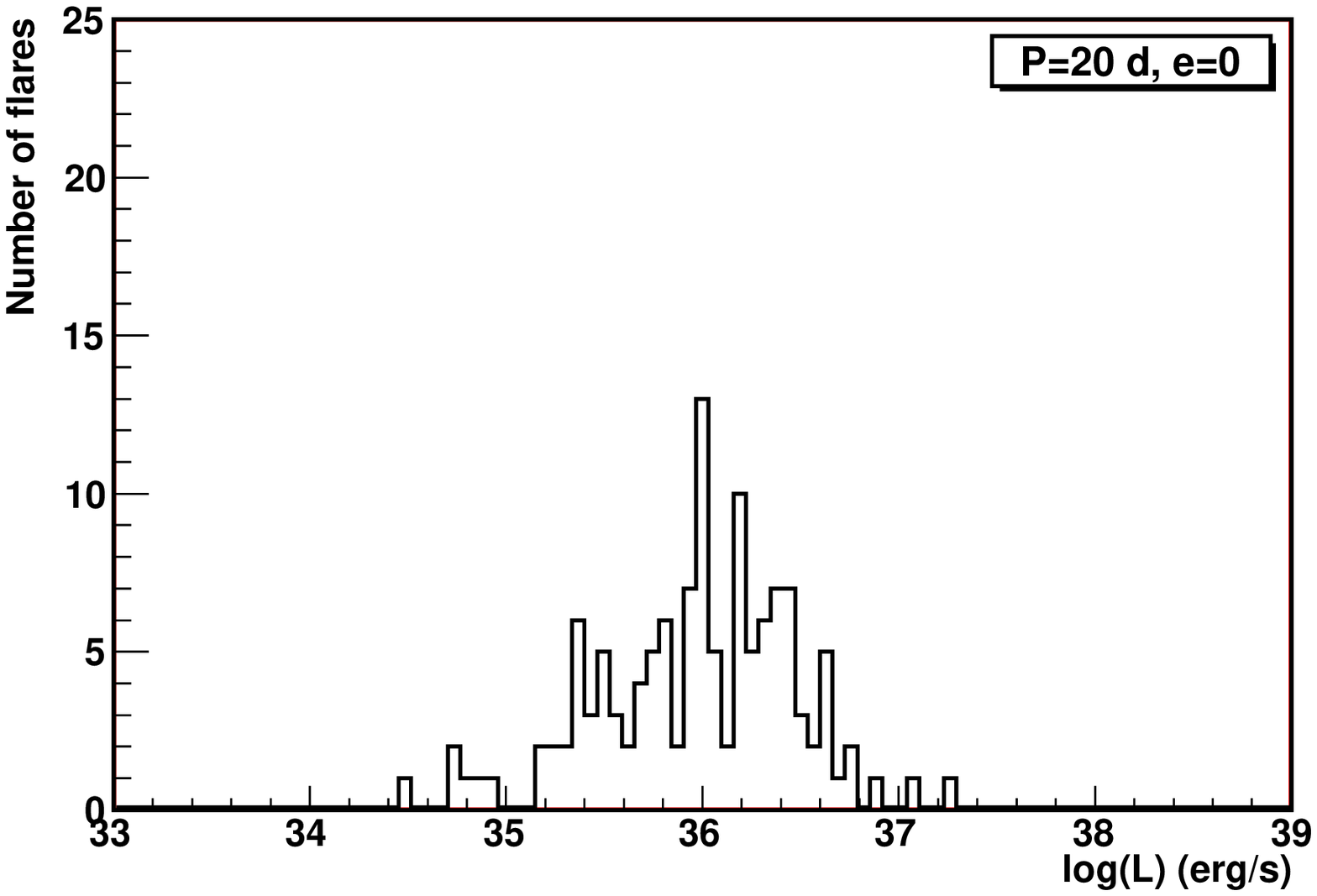} &
\includegraphics[width=8cm]{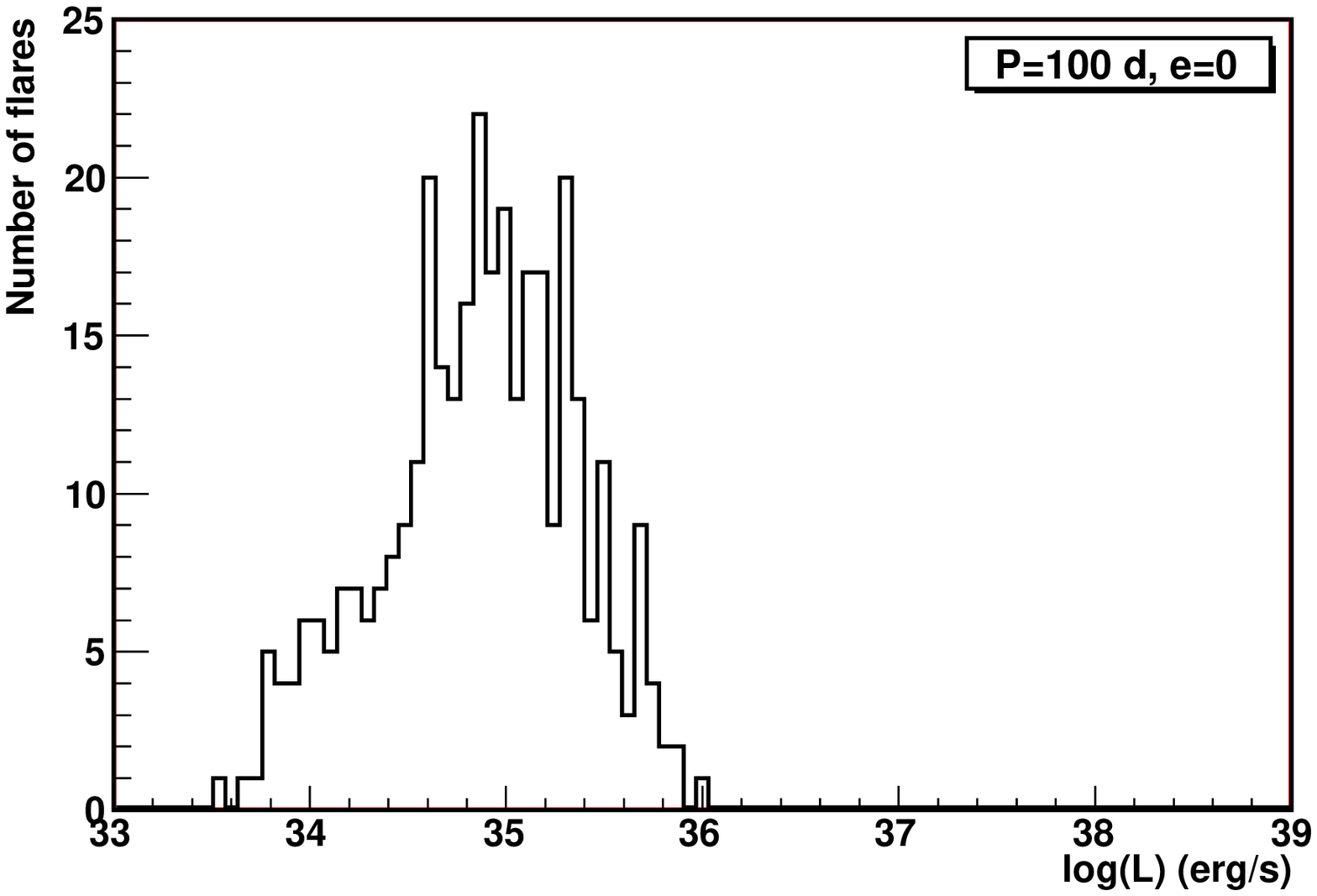} \\
\includegraphics[width=8cm]{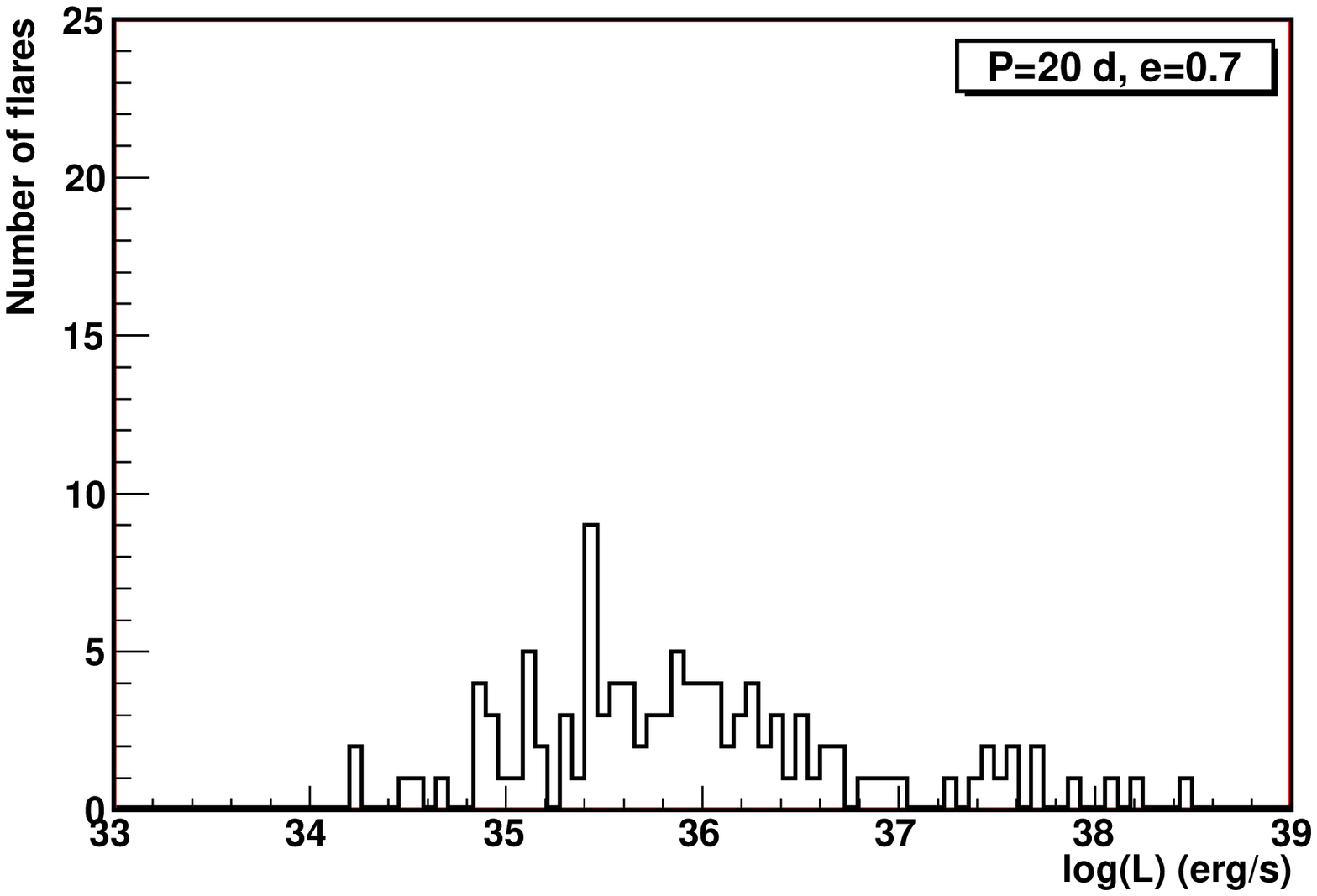} &
\includegraphics[width=8cm]{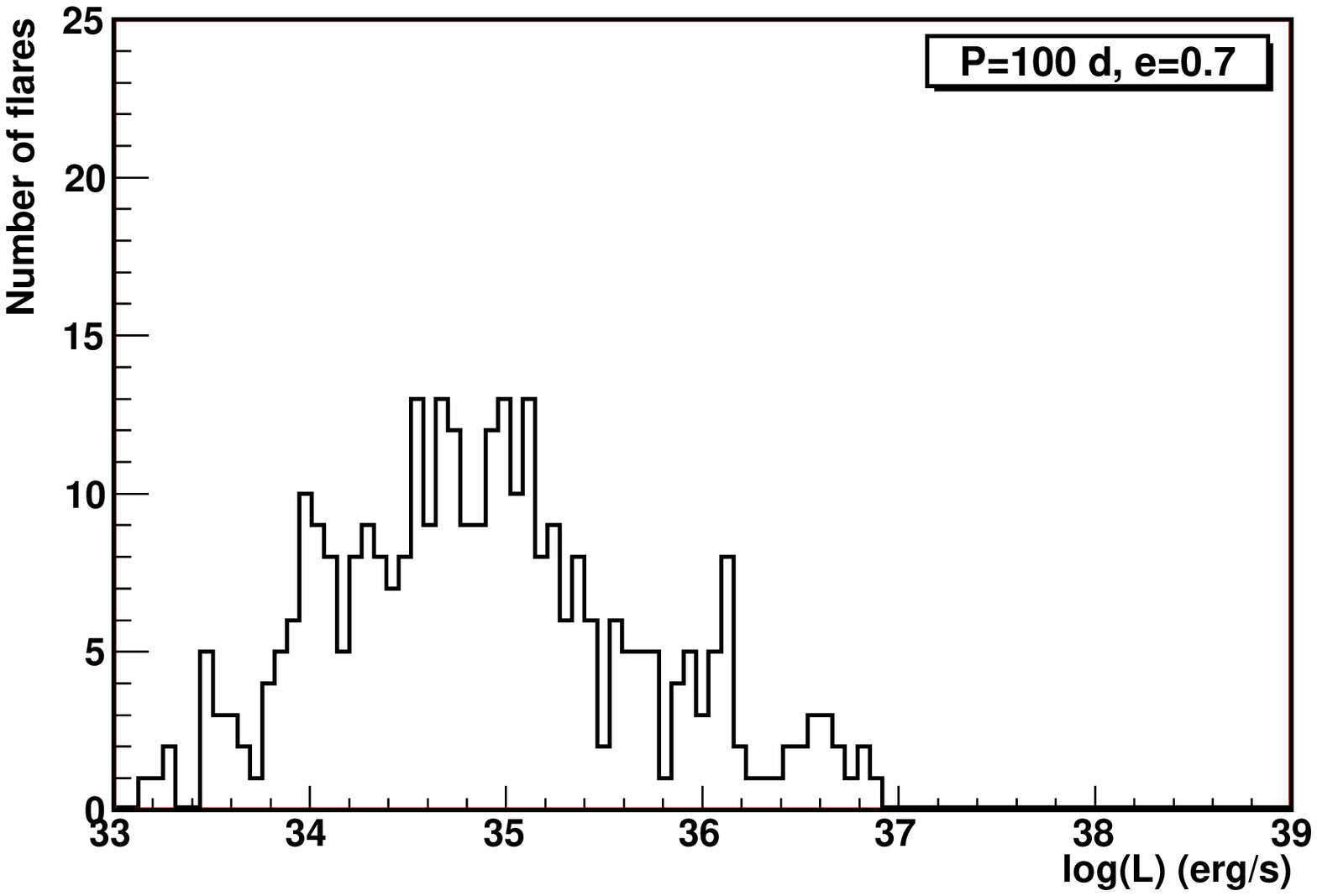} 
\end{tabular}
\end{center}
\caption{Theoretical histograms of the flare luminosities for
          different orbital periods and eccentricities of the binary
          systems. The time interval for each histogram 
          corresponds to the orbital period.}
\label{figura_istogrammi_teorici}
\end{figure}

Assuming the following parameters for the supergiant wind
$M_{OB}=30 \ M_{\odot}$, $R_{OB}=23.8 \ R_{\odot}$, 
$v_{\infty}=1700 \ km \ s^{-1}$, $\beta=0.8$, $v_0=10 \ km \ s^{-1}$,
$\zeta=2$, $\gamma=-1.5$, 
$M_a = 5 \times 10^{19} \ g$
$M_b= 10^{22} \ g$,
$\dot{M}_{cl}/\dot{M}_{wind}=0.85$,
we calculated different histograms of the flare luminosities for
different orbital periods and eccentricities of the binary systems 
(see Figure \ref{figura_istogrammi_teorici}).

\section{Comparison of observed flares with theory}

\begin{figure}[htbp]
\begin{center}
\begin{tabular}{cc}
\includegraphics[width=8cm]{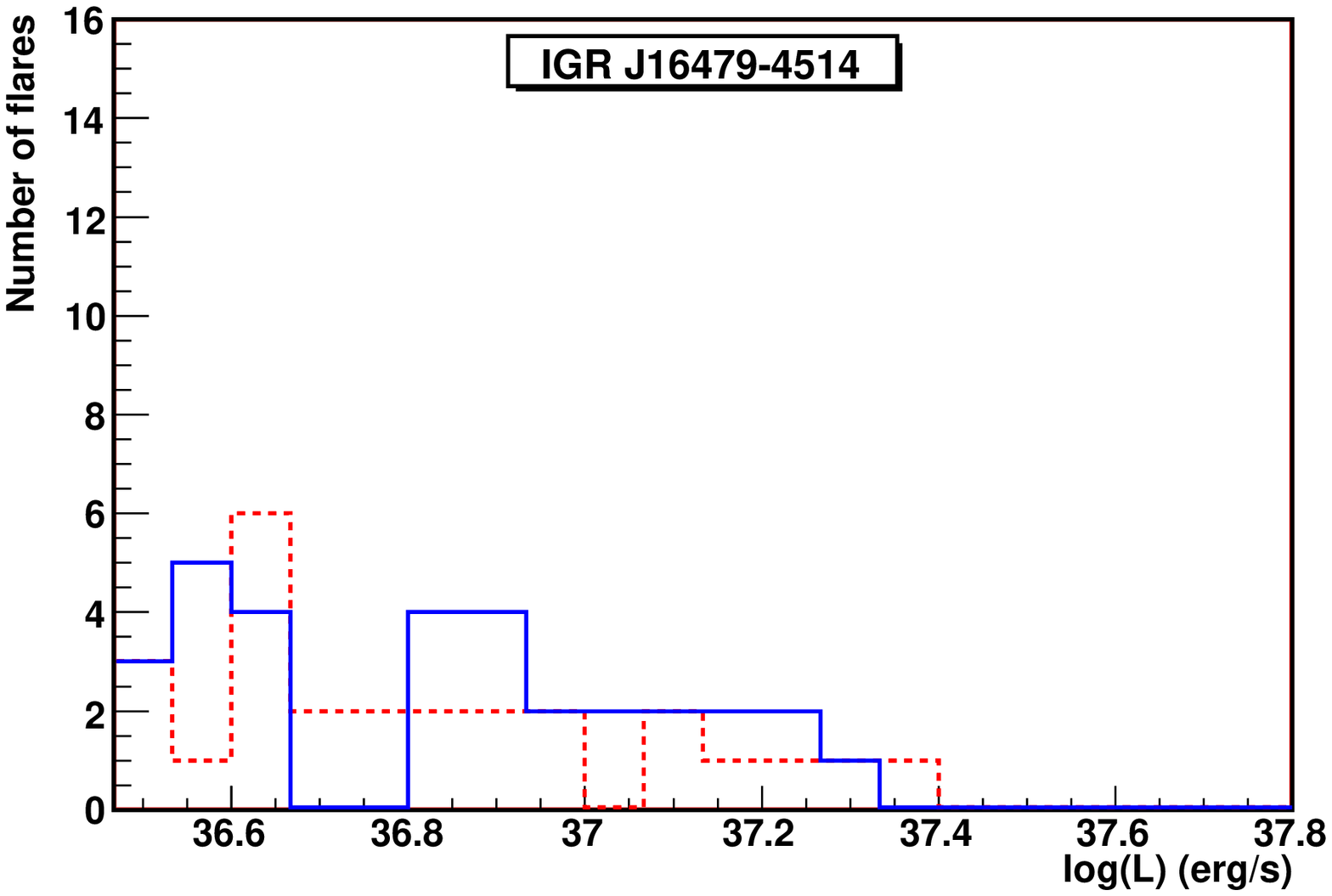} &
\includegraphics[width=8cm]{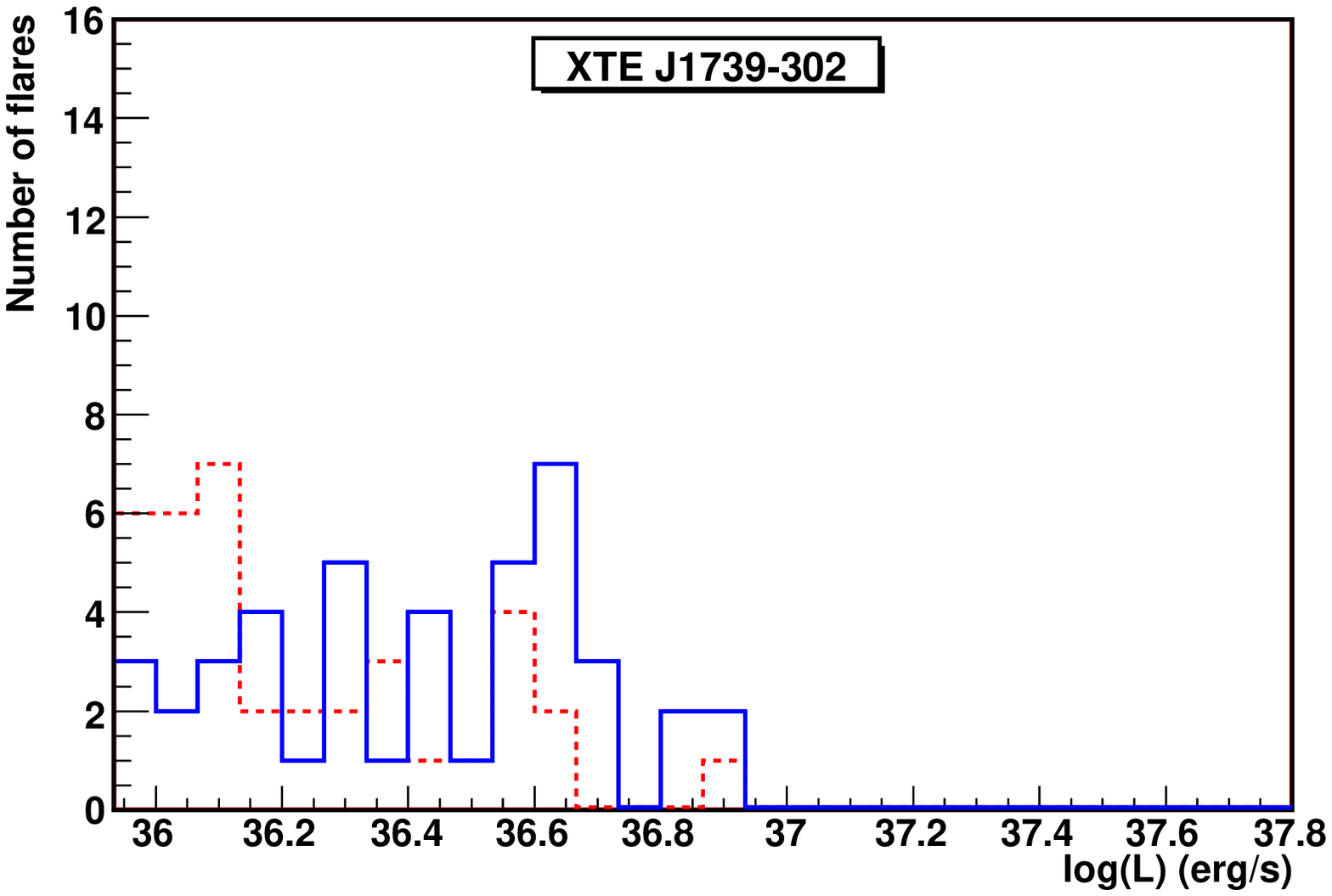} \\
$P_{orb}=30 \ d$, $e=0.4$  & $P_{orb}=70 \ d$, $e=0.4$ \\
\end{tabular}
\end{center}
\caption{Comparison of histograms of observed flares with theory for
  IGR~J16479-4514 and XTE~J1739-302. Dashed lines mark
  our calculated histograms, solid lines mark observed 
  luminosity distributions.}
\label{figura_confronto_istogrammi}
\end{figure}

We have applied our spherical clumpy wind model to the \emph{INTEGRAL}
archival observations of IGR~J16479-4514 and XTE~J1739-302,
for which we found a significant number of flares.
Our preliminary results are shown in Figure 
(\ref{figura_confronto_istogrammi}).
For these two sources we obtained 
the calculated histograms assuming the parameters 
for the supergiant winds reported in Section 
(\ref{Section Clumpy stellar winds}), 
in a time interval equal to the exposure time of 
IGR~J16479-4514 and XTE~J1739-302
(see Table \ref{tabella_parametri_spettri}).
With the spectral parameters of Table (\ref{tabella_parametri_spettri})
we obtained the flare luminosities in the energy 
range $1-200$~keV, for a distance $d=4.9 \ kpc$ 
for IGR~J16479-4514, and a distance $d=2.7 \ kpc$ 
for XTE~J1739-302 \cite{Rahoui et al. 2008}.
The comparison between observed distributions and the calculated
histograms reported in Figure (\ref{figura_confronto_istogrammi})
is limited by the fact that \emph{INTEGRAL} can detect only the 
``high luminosity tail'' of the flare distribution 
(see Figures \ref{figura_istogrammi_teorici} 
and \ref{figura_confronto_istogrammi}).

The flare luminosity distributions 
are well reproduced with
$P_{orb}=30 \ d$ and $e=0.4$ for IGR~J16479-4514, 
and $P_{orb}=70 \ d$ and $e=0.4$ for XTE~J1739-302. 
With the Kolmogorov-Smirnov test we found  
a probability of $70.6\%$
that the observed and calculated histograms 
for IGR~J16479-4514
have the same distribution.
In the case of XTE~J1739-302 this probability is $7.3\%$.
Moreover, the observed flare duration is well reproduced by the
calculated flare duration ($\sim 10^3 \ s$).
The low probability obtained with the
Kolmogorov-Smirnov test for XTE~J1739-302
indicates that for this source we need an higher number
of flares for the comparison between the
calculated histogram and the flare luminosity distribution. 
We note that the estimated orbital period 
for XTE~J1739-302 is different from that
obtained by Blay et al. (2008) \cite{Blay et al. 2008}
($P_{orb} \approx 8 \ d$), and that an
orbital period of $\approx 70 \ d$ with an eccentricity of $0.4$ could
explain the absence of flares during the whole Key Programme 2
period, lasting 32 days, reported by Blay et al. (2008).

In this framework we can study the properties of the supergiant winds
of the SFXTs by means of observable
flare luminosities, flare frequency, flare durations, and quiescent
level of the source.

These are preliminary results on the comparison between 
the spherical clumpy wind model and X-ray observations. 
Our future work will deal with anisotropic clumpy winds, where there is a
denser clumpy wind component in the form of an equatorial wind disk
around the supergiant star, and a polar spherically symmetric clumpy
wind component, as proposed by Sidoli et al. (2007) 
\cite{Sidoli et al. 2007}. We will also
compare this model with the data obtained from the still on-going
monitoring with \emph{Swift} of these 4 SFXTs \cite{Sidoli et al. 2008};
\cite{Romano et al. 2008b}.

\begin{acknowledgments}
LS, AP and SM acknowledge the Italian Space Agency financial
and programmatic support via contract I/008/07/0.
\end{acknowledgments}


\begin{thebibliography}{99}
   \bibitem{Abbott 1982}  Abbott, D. C., 
               1982,                          
               \emph{The Astrophysical Journal}
                \textbf{259}, 282    
   \bibitem{Castor et al. 1975}  Castor, J. I., 
               Abbott, D. C., 
               Klein, R. I., 
               1975,                          
               \emph{The Astrophysical Journal}
                \textbf{195}, 157
   \bibitem{Blay et al. 2008}  Blay, P.,
                         et al.
                         2008,                          
                         \emph{	arXiv:0806.4097v1}
   \bibitem{Bozzo et al. 2008}  Bozzo, E.,
                         Falanga, M.,
                         Stella, L.,
                         2008,                          
                         \emph{arXiv:0805.1849v1}
   \bibitem{Howk et al. 2000}  Howk J.C.,
                               Cassinelli J.P.,
                               et al., 
                               2000,                          
                               \emph{Astrophysical Journal}
                               \textbf{534}, 348                
\bibitem{in't Zand 2005}  in 't Zand J.J.M., 
               2005,                          
               \emph{A\&A}
                \textbf{441}, L1
   \bibitem{Kudritzki et al. 1989}  Kudtritzki, R. P.,
                                     Pauldrach, A.,
                                     Puls, J.,
                                     Abbott, D. C., 
                                     1989,                          
                                     \emph{A\&A}
                                     \textbf{219}, 205 
   \bibitem{Lamers and Cassinelli 1999}  Lamers, H.J.G.L.M,
                                          Cassinelli J.P.,
                                          1999,                          
                                          \emph{Introduction to the
                                            Stellar Winds, Cambridge
                                            University Press}
   \bibitem{Lepine and Moffat 2008}  Lepine, S.,
                                     Moffat, A. F. J.,
                                     2008,                          
                                     \emph{arXiv:0805.1864v1 }
   \bibitem{Lucy and White 1980}  Lucy, L. B.,
                                  White, R. L.,
                                  1980,                          
                                  \emph{The Astrophysical Journal}
                                  \textbf{241}, 300
   \bibitem{Negueruela et al. 2006}  Negueruela, I.,
                                     Smith, D.M.,
                                     Reig, P.,
                                     Chaty, S.,
                                     Torrej\'on, J.M.
                                     2006,   
                                     \emph{Proceedings of the ``The X-ray Universe 2005'', 26-30 September 2005, El
                                       Escorial, Madrid, Spain}
                                     \textbf{ESA SP-604, ed. A. Wilson}, 165

   \bibitem{Rahoui et al. 2008}  Rahoui, F., 
               Chaty, S., 
               Lagage, P. O.,
               Pantin, E.,
               2008,                          
               \emph{arXiv:0802.1770v1} 
\bibitem{Romano et al. 2008a} Romano, P., 
                              Sidoli, L.,
                              Cusumano, G.,
                              Evans, P. A.,
                              Ducci, L.,
                              Krimm, H. A.,
                              Vercellone, S.,
                              Page, K. L.,
                              Beardmore, A. P.,
                              Burrows, D. N.,
                              Kennea, J. A.,
                              Gehrels, N.,
                              La Parola, V.,
                              Mangano, V.,
                              2008a,                          
                              \emph{arXiv:0810.1180}
\bibitem{Romano et al. 2008b}  Romano, P., 
                              Sidoli, L.,
                              Mangano, V.,
                              Vercellone, S.,
                              Kennea, J. A.,
                              Cusumano, G.,
                              Krimm, H. A.,
                              Burrows, D. N.,
                              Gehrels, N.,
                              2008b,                          
                              \emph{The Astrophysical Journal}
                              \textbf{680}, L137
   \bibitem{Runacres and Owocki 2005}  Runacres, M. C., 
                                       Owocki, S. P., 
                                       2005,                          
                                       \emph{A\&A}
                                       \textbf{429}, 323
   \bibitem{Sguera et al. 2005}  Sguera V., 
                                 Barlow E.J., 
                                 Bird A.J.,
                                 et al.,
                                 2005,                          
                                 \emph{A\&A}
                                 \textbf{444}, 221    
   \bibitem{Shimada et al. 1994}  Shimada, M. R.,
               Ito, M.,
               Hirata, R.
               Horaguchi, T., 
               1994,                          
               \emph{IAUS}
                \textbf{162}, 487
   \bibitem{Sidoli et al. 2007}  Sidoli, L.,
                                 Romano, P.,
                                 Mereghetti, s.,
                                 Paizis, A.,
                                 Vercellone S.,
                                 Mangano, V.,
                                 Vercellone, S.,
                                 G\"{o}tz, D.,
                                 2007,                          
                                 \emph{A\&A}
                                 \textbf{476}, 1307
   \bibitem{Sidoli et al. 2008}  Sidoli, L.,
                                 Romano, P.,
                                 Mangano, V.,
                                 Pellizzoni, A.,
                                 Kennea, J. A.,
                                 Cusumano, G.,
                                 Vercellone, S.,
                                 Paizis, A.,
                                 Burrows, D. N.,
                                 Gehrels, N.,
                                 2008,                          
                                 \emph{ApJ in press,
                                   arXiv:0805.1808v1}
   \bibitem{Sidoli 2008}         Sidoli, L.,
                                  2008,                          
                                 \emph{arXiv:0809.3157v1}

\end{thebibliography}
\end{document}